# Properly Defining Erlang Lists

## Typing Erlang Lists with Set-Theoretic Types

Albert Schimpf, RPTU Kaiserslautern-Landau, Germany
Stefan Wehr, Offenburg University of Applied Sciences, Germany
Annette Bieniusa, RPTU Kaiserslautern-Landau, Germany

- define list types, proper and improper lists
- encoding of proper and improper lists as recursive types of cons cells implemented in etylizer
- found one OTP compiler bug
- suggest a one-line sound Dialyzer improvement related to lists (finds 6 dead-code in OTP code base)
- comparison of beam VM checkers
- catalogue how other languages design list and cons cells

Additional Key Words and Phrases: set-theoretic types, semantic subtyping, Erlang, lists, improper lists, type checking

## 1 Introduction

Lists and tuples are the most common compound data structures in Erlang. A list is either the empty list `[]` or a cons cell `[H | T]`, and the tail `T` may be any term:

```
[1, 2, 3]                  % proper:   [1 | [2 | [3 | []]]]
[1 | foo]                  % improper: the last tail is foo
[<<"ab">>, $c | <<"d">>]   % an iolist: a binary may end the chain
```

A list whose chain of tails ends in `[]` is *proper*; otherwise it is *improper*. The guard `is_list([1 | foo])` succeeds, because it inspects only the outermost constructor, but `length([1 | foo])` raises `badarg`. Moreover, `list_to_binary([2 | <<1>>])` returns `<<2,1>>`, while `list_to_binary(<<1>>)` raises `badarg`.

Erlang's type language describes lists by an element type and a termination type, through `list(T)`, `nonempty_list(T)`, `maybe_improper_list(T, U)` and `nonempty_improper_list(T, U)`, and defines `iolist()` in terms of them ("Types and Function Specifications" 2025; Lindahl and Sagonas 2007).

A type checker for Erlang's list type has to settle two questions.

The first is how to represent list types so that subtyping is decided correctly, and as precisely as the checker needs. A checker must reject

```
-spec widen(list(a | b)) -> list(a) | list(b).
widen(L) -> L.
```

because `widen([a, b])` returns a list that mixes `a` and `b` and is therefore neither a `list(a)` nor a `list(b)`. How much precision is needed depends on what the checker does with the types. Knowing what the elements of a list can be and how the list ends is enough for a checker that does not look for non-exhaustive or redundant clauses. A checker that does could notice that the third clause of

```
case L of [] -> 1; [1 | _] -> 2; [1, 2 | _] -> 3; _ -> 4 end
```

can never match. What it must not do is report the third clause of

```
case S of "" -> 1; "trace" -> 2; "trace2" -> 3; _ -> 4 end
```

which does match the string `"trace2"`. A type that only says that a string pattern is a list of characters cannot tell the second and the third pattern apart. The pattern typing must approximate, the

Authors' contact information: Albert Schimpf, RPTU Kaiserslautern-Landau, Kaiserslautern, Germany, schimpf@cs.uni-kl.de; Stefan Wehr, Offenburg University of Applied Sciences, Offenburg, Germany, stefan.wehr@hs-offenburg.de; Annette Bieniusa, RPTU Kaiserslautern-Landau, Kaiserslautern, Germany, bieniusa@cs.uni-kl.de.

second clause then seems to match every remaining string, and the third clause therefore seems dead. Avoiding such false alarms requires types that know precisely which element is where.

The second question is what to make of improper lists. Improper lists are rarely used. In the source code of OTP, fewer than 0.2% of the list literals are explicitly improper, and in three applications we measured the only improper list types in specifications are `iolist()` and `iodata()`. A checker may decide to ignore them and accept the unsoundness, as eqWAlizer does (todo ref). Still, the list representations may intermingle in unexpected ways. The syntax for lists is also the syntax for improper lists. Consequently, they are easily created by accident. One character separates `[X, Y]` from `[X | Y]`, and `L ++ X` builds an improper list whenever `L` is non-empty and `X` is not a list. Once created, they reach code that never mentions them.

The operator `++` can even return a value that is not a list: `[] ++ foo` is `foo`, although the reference manual says that `++` "returns the resulting list". OTP's own compiler overlooks this case. Its type for `++` assumes a proper list, and OTP 28.5 compiles a function that tests the tail of such a result incorrectly (Section 2.2).

Erlang's standard library has an example of code that unintuitively mixes proper and improper lists. The function `is_set/1` of the `ordsets` module checks whether a term is an ordered set:

```
-spec is_set(Ordset) -> boolean() when
      Ordset :: term().
is_set([E|Es]) -> is_set(Es, E);
is_set([]) -> true;
is_set(_) -> false.

is_set([E2|Es], E1) when E1 < E2 ->
    is_set(Es, E2);
is_set([_|_], _) -> false;
is_set([], _) -> true.
```

The specification promises a boolean for every term, but `ordsets:is_set([1 | foo])` crashes with `function_clause`: the first clause passes the tail `foo` to the helper `is_set/2`, whose clauses cover only cons cells and `[]`. A checker finds this bug only if it knows that the tail of a list matched against `term()` can be any term. For that it has to know what the list types that mention improper lists mean, namely whether `[]` is a `maybe_improper_list(integer(), atom())`, and whether a bare `foo` is. The reference manual does not say. It offers three types for improper lists, `maybe_improper_list`, `nonempty_improper_list` and `nonempty_maybe_improper_list`, and says little more about them than that their arguments are the "contents" and the "termination" and that they "define the set of Erlang terms one would expect" ("Types and Function Specifications" 2025). It does not say how the three differ, and two of them turn out to be names for the same type (Section 2). The developers of Gradualizer found three incompatible readings among themselves and left the question open (Norell 2018). Similarly, Söderqvist (2020) tried to recover the meaning from the output of TypEr. A user reported the definition of `iolist()` as a bug to both Elixir and OTP, reading the "maybe" in `maybe_improper_list` as already allowing `[]`, and the maintainers who answered explained the name in different ways (Qqwy 2022a; 2022b). The answers to a long-standing Stack Overflow question disagree on whether an improper list is a list at all (Stack Overflow 2009). Finally, etylizer itself encoded the types wrongly twice, once without `[]` and once with the bare termination value (Schimpf 2026) (Appendix A.7).

The first question is well-defined. "What is a proper list?" and what `list(T)` means is not in dispute. Thus, a checker should have no difficulty with this part. We will see that a correct answer takes more care in a set-theoretic type system (Section 3 and Appendix A).

As for the second question, here it is a matter of interpretation. There is no document to be wrong against, and every project had to decide for itself what the improper list types mean.

We first settle what improper lists and their types mean (Section 2). We then show how to implement lists, proper and improper, correctly in a set-theoretic type system (Section 3), using etylizer (Schimpf et al. 2022; 2026), a set-theoretic type checker for Erlang. Finally, we look at other languages and type systems that have similar constructs (Section 4). The appendix records etylizer's earlier approaches and why they failed (Appendix A).

## 2 Proper and Improper Erlang Lists

Before defining Erlang lists and what their types mean, we first collect what the documentation and the OTP sources say.

### 2.1 OTP documentation and semantics

*Documentation.* The reference manual ("Types and Function Specifications" 2025) gives the grammar and the built-in type:

```
List :: list(Type)                           %% Proper list ([]-terminated)
      | maybe_improper_list(Type1, Type2)    %% Type1=contents, Type2=termination
      | nonempty_improper_list(Type1, Type2) %% Type1 and Type2 as above
      | nonempty_list(Type)                  %% Proper non-empty list

iolist() :: maybe_improper_list(byte() | binary() | iolist(), binary() | [])
```

It lists `nonempty_improper_list(Type1, Type2)` and `nonempty_maybe_improper_list(Type1, Type2)` among types "expected to be rarely used", and says that they "define the set of Erlang terms one would expect". EEP 8 (Lindahl and Sagonas 2007), which introduced the type language, uses the same words. Neither says whether `[]` belongs to `maybe_improper_list(T, U)`, whether a bare termination value does, or how `nonempty_improper_list` differs from `nonempty_maybe_improper_list`. The documentation of the `erlang` module (Ericsson AB 2026a) (Ericsson AB 2026b) adds one sentence per type. `maybe_improper_list(T, U)` is a list "that is not guaranteed to end with a `[]`", and `nonempty_improper_list(T, U)` and `nonempty_maybe_improper_list(T, U)` are each a `maybe_improper_list(T, U)` "that contains some items". This documentation relates the types to one another, but does not say what the termination type means. Three parts of the OTP sources fix a meaning implicitly, each for its own purpose.

*Runtime system.* In the runtime system, the two places that fix what a list and an iolist are are `is_list` and `list_to_binary/1`. The guard `is_list` is a tag test that succeeds for any cons cell and for `[]`. The traversal behind `list_to_binary/1` accepts `[]` or a cons cell at the top and raises `badarg` for anything else, including a bare binary. Along the chain, each element must be a byte, a binary, a nested iolist or `[]`, and the final tail must be a binary or `[]`. `iolist_to_binary/1` accepts a bare binary only because it tests for one before the traversal because it takes `iodata()`.

*Dialyzer.* Dialyzer's type representation describes a list by its contents, its termination and whether it may be empty. Its meet of a list type with `[]` is the meet of the termination with `[]`. Therefore, `[]` belongs to `maybe_improper_list(T, U)` exactly when it belongs to `U`. Non-list values are never lists. Accordingly, a bare termination value is excluded. `nonempty_improper_list(T, U)` and `nonempty_maybe_improper_list(T, U)` are both translated to the cons type `t_cons(T, U)`. The two names differ only when a type is printed, where `nonempty_improper_list` is chosen if `U` excludes `[]`. Two disabled assertions record the intended reading. The constructor for `maybe_improper_list(T, U)` would check that `U` contains `[]` ("would be nice to have but does not work with remote types"), and a commented-out constructor for improper lists would check that `U` does not. In Dialyzer's vocabulary, "maybe" thus refers to the termination. A maybe-improper list may end in `[]` and so may be proper. For the guard `is_list(X)`, Dialyzer uses `maybe_improper_list()` as the over-approximation of the values that pass and `list()` as the under-approximation.

*Compiler.* The compiler's type lattice has two list types, one for non-empty lists and one for lists that may be empty. Both record an element type and a terminator, which the source describes as "the tail of the last cons cell". For proper lists, the terminator is the empty list. The type for lists that may be empty is the non-empty type together with the empty list. In particular, it contains `[]` even if its terminator does not. For the guard `is_list(X)`, the compiler uses this type with any element type and any terminator.

Table 1: Membership according to the OTP sources and to etylizer. MIL and NEIL abbreviate `maybe_improper_list` and `nonempty_improper_list`. Runtime: the result of `list_to_binary/1` and `is_list/1`; Dialyzer (OTP 28.5) and etylizer (current revision): whether a function returning the value is accepted with the type as its specification; compiler: read from the type lattice, with MIL$(T, U)$ as its type for lists that may be empty, with element type `T` and terminator `U`. $^w$: accepted with a warning that the cons produces an improper list. Bold: the OTP sources disagree.

| | Runtime | Dialyzer | Compiler | etylizer |
|---|---|---|---|---|
| `foo` ∈ MIL(`integer(),atom()`) | — | no | no | no |
| `[]` ∈ MIL(`integer(),atom()`) | — | **no** | **yes** | yes |
| `[]` ∈ MIL(`integer(),atom() \| []`) | — | yes | yes | yes |
| `[1 \| foo]` ∈ NEIL(`integer(),atom()`) | — | yes$^w$ | yes | yes |
| `[]` ∈ NEIL(`integer(),atom()`) | — | no | no | no |
| `<<"x">>` ∈ `iolist()` | no | no | — | no |
| `[2 \| <<1>>]` ∈ `iolist()` | yes | yes$^w$ | — | yes |
| `is_list(X)` establishes | cons or `[]` | MIL() | MIL() | MIL() |

*Summary.* Table 1 collects the answers. Every source that takes a position agrees that a bare termination value is not a list: a list is a cons cell or `[]`, and the termination type describes the tail of the last cons cell. The Common Lisp standard says the same about dotted lists (Section 4). The sources disagree on the empty list. Dialyzer treats `[]` as a list whose termination is `[]` itself, the compiler always includes it. These two readings coincide whenever the termination type contains `[]` (which holds for `iolist()` and for all predefined aliases).

## 2.2 Concatenation

The list concatenation operator `++` "appends its second argument to its first and returns the resulting list", in the words of the reference manual (Ericsson AB 2026c). For example, `[1, 2] ++ [3]` is `[1, 2, 3]`. Its specification is `'++'(list(), term()) -> term()`. It is typically used to append one list to another and, since strings are lists of characters, to concatenate strings, as in `"foo" ++ Name`.

The result of the operator `++` need not be a list, however, and OTP describes the operator in several places. The operator replaces the `[]` at the end of its first argument by its second argument, which it never inspects (Table 2). The first argument must be a proper list and the second may be any term. If the first argument is `[]`, the result is the second argument itself, list or not. Consequently, the sentence in the manual is wrong for `[] ++ foo`. For a first argument of type `list(T)` and a second of type `U`, the result has type `U | nonempty_improper_list(T, U)`. Unless `U` is a list type, this type contains values that are not lists. This is what `maybe_improper_list(T, U)` would be if a bare termination value were a list. It is the type of a result of `++`, not a list type. However, etylizer's first cons encoding used this set for `maybe_improper_list(T, U)` (Appendix A.7).

*Specifications.* OTP specifies the operation three times.[1] The specification of the operator itself is true and says nothing about the result. Its alias `erlang:append/2` is specified to return

Table 2: What `L ++ X` returns. The operator raises `badarg` only if `L` is not a proper list.

| `L` | `X` | `L ++ X` | The result is |
|---|---|---|---|
| `[1, 2]` | `[3]` | `[1, 2, 3]` | a proper list |
| `[1, 2]` | `foo` | `[1, 2 \| foo]` | an improper list |
| `[]` | `[3]` | `[3]` | a proper list |
| `[]` | `foo` | `foo` | not a list |

[1] `erts/preloaded/src/erlang.erl` (line 13271 and lines 6866–6868) and `lib/stdlib/src/lists.erl` (lines 226–232), in Erlang/OTP 28.5.

`maybe_improper_list()` for a second argument of type `term()`. Since `erlang:append([], foo)` returns `foo`, this holds only if every term is a `maybe_improper_list()`, the reading that the runtime system, Dialyzer and the compiler reject (Section 2.1). And `lists:append/2`, which is defined as `L1 ++ L2`, is specified for proper lists only. Both arguments and the result have type `[T]`.

*Dialyzer.* Dialyzer does not use these specifications. Instead, it has a type of its own for the operator.[2] An empty first argument gives the second argument, "even if X2 is not a list", a non-empty one gives a non-empty list that ends in the second argument, and a possibly empty one gives the union of both, which is the type stated above. Dialyzer's warning about improper lists does not cover the operator, however. It reports that `[1 | foo]` "will produce an improper list" and says nothing about `[1] ++ foo`.

*Compiler.* The compiler gets the operator wrong. Its source notes that "`[] ++ RHS` yields `RHS`, even if `RHS` is not a list", but its type for the operator takes the list part of the result to be proper,[3] and the optimiser relies on that type. For

```
f(L, X) when is_list(L), is_atom(X) ->
    case L ++ X of [_ | T] -> is_atom(T); _ -> other end.
```

the call `f([1], foo)` must return `true`, because `[1] ++ foo` is `[1 | foo]` and its tail is the atom `foo`. Compiled with OTP 28.5, it returns `false`, because the compiler wrongly concludes that the tail `T` is a list and removes the test. Without type-based optimisations (`+no_type_opt`), the result is `true`. OTP's development branch of October 2026 (commit `cfe4f03`) gives the same results.

*etylizer.* etylizer's type for `++` (Section 3.3) is that of `lists:append/2`, which requires proper lists on both sides. etylizer therefore rejects the second and the fourth row of Table 2 (which is sound but turns away programs that run). The type that covers all four rows is the intersection of two arrows, `([], b) -> b` and `(nonempty_list(a), b) -> nonempty_improper_list(a, b)`.

### 2.3 The guard `is_list/1`

Dialyzer uses two types for the guard `is_list(X)` (Section 2.1), namely `maybe_improper_list()` for the values that pass and `list()` when it subtracts what a clause handles from the clauses after it. Using `list()` for later clauses is sound, since every proper list passes `is_list/1`, but it costs precision, and one line of Dialyzer's source repairs it.[4] In Dialyzer's type operations, `maybe_improper_list()` minus `list()` is `maybe_improper_list()` again. Consequently, a clause guarded by `is_list(X)` removes nothing from an argument that is known only to be a list. In OTP's ASN.1 compiler, for example, the second clause of

```
check_choice(S,Type,Components) when is_list(Components) -> ...;
check_choice(_S,_,[]) -> [].
```

can never match, because `[]` passes `is_list/1`. However, Dialyzer currently does not report it. If one replaces `list()` by `maybe_improper_list()` in that line, the changed Dialyzer then reports 6 dead clauses across 5 modules. This change keeps the approximation sound, because `is_list/1` succeeds for exactly the values of `maybe_improper_list()`.

### 2.4 A proposal for the reference manual

The reference manual can state the semantics of list types in a few sentences. We propose the following text.

> A *list* is either the empty list `[]` or a cons cell `[H | T]`, where `T` is any term; `is_list/1` returns `true` exactly for these terms. A list is *proper* if its chain of tails ends in `[]`, and *improper* otherwise. For

---

[2] `lib/dialyzer/src/erl_bif_types.erl`, lines 410–426.

[3] `lib/compiler/src/beam_call_types.erl`, lines 526–530.

[4] `lib/dialyzer/src/dialyzer_typesig.erl`, line 1040.

Table 3: The four list types under the proposed rule.

| | ends in `[]` (proper) | ends in `U` |
|---|---|---|
| may be `[]` | `list(T)` | `maybe_improper_list(T, U)` |
| non-empty | `nonempty_list(T)` | `nonempty_improper_list(T, U)` |

types `T` and `U`, a *chain over `T` ending in `U`* is a term `[V1, ..., Vn | W]` with $n \geq 1$, every `Vi` of type `T`, and `W` of type `U`.

- `nonempty_improper_list(T, U)` and `nonempty_maybe_improper_list(T, U)` denote the chains over `T` ending in `U`; the two names are synonyms.
- `maybe_improper_list(T, U)` denotes `[]` and the chains over `T` ending in `U`.
- `nonempty_list(T)` is `nonempty_improper_list(T, [])`, and `list(T)` is `maybe_improper_list(T, [])`.

A value of type `U` does not belong to `maybe_improper_list(T, U)` unless it is `[]` or itself such a chain; in particular, a binary is an `iodata()` but not an `iolist()`. After the guard `is_list(X)` succeeds, `X` has type `maybe_improper_list()`, not `list()`.

Here, *nonempty* decides whether `[]` belongs to the type, and *improper* whether the chain may end in something other than `[]`. The four list types form the grid of Table 3, and the proper types are the improper ones with `U = []`. Under Dialyzer's rule, `maybe_improper_list(T, U)` and `nonempty_improper_list(T, U)` coincide whenever `U` excludes `[]`. The grid then collapses for such `U`. The proposed rule is the one of the compiler, of the runtime's `is_list/1`, of Elixir and etylizer, and of eqWAlizer, which otherwise ignores `U`. A member of the OTP team has stated it in the same words, in answers to the bug reports mentioned in Section 1: "`maybe_improper_list(X, Y)` is the union of `[]` | `nonempty_improper_list(X, Y)`" (Qqwy 2022a), and "it's `list` and `nonempty_list` that are syntactic sugar for `maybe_improper_list(X, [])` and `nonempty_improper_list(X, [])`, respectively" (Qqwy 2022b). This rule differs from Dialyzer and Gradualizer only when `U` excludes `[]`, a case that Dialyzer's own disabled assertion says should not occur. Since Dialyzer mixes both readings in its type operations, it needs changes under either rule. An alternative is to turn that assertion into a rule of the language and require `U` to contain `[]`. Both readings then coincide by construction, but specifications that are accepted today, such as `maybe_improper_list(integer(), atom())`, would become errors. We prefer to give every specification a meaning.

The proposal changes no program and no runtime behaviour, only the documentation and the checkers that disagree with it. Like EEP 71, which redefined parts of the type language and likewise changed no runtime behaviour (Högberg and Klyuchnikov 2024), it could be submitted as an Erlang Enhancement Proposal.

## 3 Lists in a Set-Theoretic Type System

This section shows how to represent lists, proper and improper, so that subtyping is decided correctly and precisely, and what etylizer builds on the representation. In a set-theoretic type system, each type $t$ denotes a set of values $[\![t]\!]$, subtyping $s \leq t$ is set containment, and types are closed under union $\vee$, intersection $\wedge$, negation $\neg$ and recursion $\mu X.\, t$ (Frisch et al. 2008).

### 3.1 Lists as cons cells

Let $\mathtt{cons}(h, t)$ denote the cons cells whose head is in $h$ and whose tail is in $t$:

$$[\![\mathtt{cons}(h, t)]\!] = \{\, [v \mid w] \mid v \in [\![h]\!],\ w \in [\![t]\!] \,\}.$$

A cons cell decomposes uniquely into head and tail. Therefore, `cons` is a Cartesian product up to a bijection, and the standard emptiness check for products (Frisch et al. 2008) is exact for it. Proper list types are recursive types over cons cells:

$$\texttt{list}(t) = \mu X.\ \texttt{[]} \vee \texttt{cons}(t, X)$$
$$\texttt{nonempty_list}(t) = \texttt{cons}(t, \texttt{list}(t))$$

These definitions are the standard encoding of lists as recursive types over a product, which CDuce uses for its sequences (Section 4.1). Values have exact types: `[a, b]` has type $\texttt{cons}(\texttt{a}, \texttt{cons}(\texttt{b}, \texttt{[]}))$, and the string `"foo"` has type

$$\texttt{cons}(\texttt{\$f}, \texttt{cons}(\texttt{\$o}, \texttt{cons}(\texttt{\$o}, \texttt{[]}))).$$

### 3.2 Improper lists

With cons cells, the definition of Section 2.4 translates directly. etylizer encodes the improper list types as follows:

$$\texttt{nonempty_improper_list}(t, u) = \texttt{cons}(t, \mu X.\ u \vee \texttt{cons}(t, X))$$
$$\texttt{maybe_improper_list}(t, u) = \texttt{[]} \vee \texttt{nonempty_improper_list}(t, u)$$
$$\texttt{iolist}() = \mu X.\ \texttt{[]} \vee \texttt{nonempty_improper_list}(\texttt{byte}() \vee \texttt{binary}() \vee X,\ \ \texttt{binary}() \vee \texttt{[]})$$

As in Dialyzer, `nonempty_maybe_improper_list` means the same as `nonempty_improper_list`. The guard `is_list(X)` establishes `maybe_improper_list()`, that is, $\texttt{[]} \vee \texttt{cons}(\texttt{any}(), \texttt{any}())$. The inner type $\mu X.\, u \vee \texttt{cons}(t, X)$, which we write $\texttt{improper_list}(t, u)$, contains the chains of length zero or more over $t$ that end in $u$. The outer cons cell makes the chain non-empty, so a bare termination value is excluded, as in all OTP sources. For the empty list, etylizer follows the compiler and the runtime system's notion of a list, and contains `[]` in every `maybe_improper_list`. This choice differs from Dialyzer only on an explicit `maybe_improper_list(T, U)` whose termination excludes `[]`, the second row of Table 1.

Proper lists are, as expected, the special case of improper lists whose termination is `[]`:

$$\texttt{list}(t) = \texttt{improper_list}(t, \texttt{[]}) = \texttt{maybe_improper_list}(t, \texttt{[]})$$
$$\texttt{nonempty_list}(t) = \texttt{nonempty_improper_list}(t, \texttt{[]})$$

All four list types are thus built on one recursive type over cons cells, $\texttt{improper_list}(t, u)$, and differ only in the termination and in whether `[]` belongs to them (Table 3).

### 3.3 Expressions and patterns

etylizer generates subtyping constraints from expressions and patterns and solves them by tallying (Castagna et al. 2015). With cons types, both map directly to the type language. For an expression $\texttt{[}e_1\ \texttt{|}\ e_2\texttt{]}$ with expected type $\tau$, the constraint is $\texttt{cons}(\alpha, \beta) \leq \tau$ with $e_1 : \alpha$ and $e_2 : \beta$. Nothing forces the tail to be a proper list, so `[1 | foo]` has type $\texttt{cons}(\texttt{1}, \texttt{foo})$. A cons pattern $\texttt{[}p_1\ \texttt{|}\ p_2\texttt{]}$ has the exact type $\texttt{cons}(t_1, t_2)$, where $t_1$ and $t_2$ are the types of $p_1$ and $p_2$. Matching a value of type $\tau$ against it adds $\tau \leq \texttt{cons}(\alpha_1, \alpha_2)$ and continues with $p_1$ at $\alpha_1$ and $p_2$ at $\alpha_2$. String literals and string patterns are cons chains of character singletons.

Exact pattern types make exhaustiveness and redundancy checks precise. The third clause of the first `case` expression in Section 1 is correctly reported (“this branch never matches”), and the third clause of the second, with the pattern `"trace2"`, is not. Several later additions build on the encoding. The operator `++` has the intersection of four arrow types, so that appending to or with a non-empty list yields a non-empty list. List comprehensions have type $\texttt{list}(\beta)$ for body type $\beta$.

### 3.4 Heterogeneous lists

Cons types can also be used to describe heterogeneous lists. These are lists in which every position has its own type, as in the HLists of Kiselyov et al. (2004). They have a fixed number of elements in a fixed order, or a fixed prefix followed by any number of elements. In the loose sense of the word, a list whose elements have different types, Erlang’s `[T | U]` is heterogeneous too, but it

is the type of the lists of any length whose elements have type T or U, and does not say which element is where.

Heterogeneous lists are common in Erlang's abstract format, the representation of parsed code that the compiler, parse transforms and etylizer itself work on (Ericsson AB 2026d). A remote type `M:N(T1, ..., Tk)` is represented as `{remote_type, Anno, [M, N, [T1, ..., Tk]]}`, a list of two atoms followed by a list of types; an annotated type `A :: T` as `{ann_type, Anno, [A, T]}`; and a record type as `{type, Anno, record, [Name, F1, ..., Fk]}`, a name followed by any number of fields.

Erlang's type language currently cannot express such types. etylizer needs them to type check itself, since it works on the abstract format. Until the type language supports them, if at all, it therefore needs a construct of its own. etylizer makes the cons type available in specifications as `etylizer:cons(Head, Tail)` and uses it to describe the abstract format in its own source:

```
-type list2(T, U)     :: etylizer:cons(T, etylizer:cons(U, [])).
-type list3(T, U, V)  :: etylizer:cons(T, etylizer:cons(U, etylizer:cons(V, []))).
-type list1star(T, U) :: etylizer:cons(T, [U]).

-type ty_ann()    :: {ann_type, anno(), list2(exp_var(), ty())}.
-type ty_remote() :: {remote_type, anno(), list3(rep_atom(), rep_atom(), [ty()])}.
-type ty_record() :: {type, anno(), record, list1star(rep_atom(), ty_field())}.
```

A clause that takes such a list apart gives every element its own type. With `rep_atom()` defined as `{atom, anno(), atom()}`, the function

```
-spec name(ty_remote()) -> {atom(), atom(), non_neg_integer()}.
name({remote_type, _, [{atom, _, M}, {atom, _, N}, Args]}) -> {M, N, length(Args)}.
```

type checks, and its single clause is exhaustive. With the closest type in Erlang's syntax, `[rep_atom() | [ty()]]`, etylizer rejects it, because the clause covers only lists of length three. Moreover, with a catch-all clause, each element would still have the union type.

### 3.5 Implementation

In etylizer a type is a record with one component per kind of value: predefined constants (including `[]`), atoms, integer intervals, lists, bitstrings, tuples, functions and maps. Each component is a binary decision diagram over the atoms of its kind. The atoms of the list component are cons cells, represented as binary tuples, and its emptiness check reuses the code for tuples. Recursive types had already been supported for user-defined types, so the encoding required no new machinery.

### 3.6 A bug in the standard library

The function `is_set/1` from Section 1 has a specification, its helper `is_set/2` has none. By default etylizer gives functions without a specification the type `dynamic()`, which effectively hides the bug. Therefore, we give the helper a type and try to make both functions type check without using `dynamic()`. Table 4 shows the natural candidates on the current revision of etylizer (`2aac9a2`, 2026-10-01)[5]. With `list()` as the first argument type, the call `is_set(Es, E)` in `is_set/1` fails, because the tail `Es` of a list matched against `term()` can be any term. With `term()`, the clauses of `is_set/2` are not exhaustive. With `maybe_improper_list()`, the call fails again, because the tail of `[1 | foo]` is `foo`, which is not a list of any kind. No type for the helper works, because the function crashes on improper lists. Adding the clause `is_set(_, _) -> false` to the helper fixes the bug, and the fixed version type checks successfully with the helper type `(term(), term()) -> boolean()`. With type inference for functions without specifications (`--gradual-mode infer`), etylizer reports the call `is_set(Es, E)`.

The bug is present in the current OTP release as of writing this paper. None of the other checkers we examined reports it.

[5]https://github.com/etylizer/etylizer

Table 4: Checking `is_set/1` with different types for its helper `is_set/2`. “call”: the call `is_set(Es, E)` in `is_set/1` fails to type check; “exh.”: the clauses of `is_set/2` are not exhaustive.

| Code | Type of `is_set/2` | Verdict |
|---|---|---|
| verbatim | `(list(), term()) -> boolean()` | rej (call) |
| verbatim | `(term(), term()) -> boolean()` | rej (exh.) |
| verbatim | `(maybe_improper_list(), term()) -> boolean()` | rej (call) |
| verbatim | inferred | rej (call) |
| fixed | `(term(), term()) -> boolean()` | acc |

## 4 Other Languages and Type Systems

Erlang inherited cons cells with arbitrary tails, and the notation for them, from older languages. In Lisp (McCarthy 1960), lists are built from cons cells, and the list notation `(a b c)` abbreviates `(a . (b . (c . nil)))`. A chain may end in any atom, written `(a b . c)`. As a result, dotted and proper lists share one notation. Common Lisp calls such a chain a *dotted list* and adds: “An atom by itself is not a dotted list” (Pitman 2005a), the same answer the OTP sources give for a bare termination value. Scheme and Racket keep the notation. For example, `(cons 1 2)` prints as `'(1 . 2)` (Flatt and PLT 2026). Prolog writes lists as `[H|T]` and allows any term as the tail (so `[a|b]` is a valid term). Erlang, which began as an interpreter written in Prolog (Armstrong 2007), uses the same notation. Elixir runs on the BEAM and shares Erlang’s lists. The Elixir documentation notes that improper lists are “generally avoided” but used for iodata and chardata (The Elixir Team 2026). Statically typed functional languages, including Gleam on the BEAM (The Gleam Team 2026), have only proper lists, because the tail of a cons must be a list of the same type.

Table 5 compares the vocabularies. In Common Lisp, *list* covers every chain of conses, and proper and dotted lists are its two finite kinds (Pitman 2005b). For example, `(listp (cons 1 2))` is true. In Scheme, “a chain of pairs not ending in the empty list is called an improper list. Note that an improper list is not a list” (Kelsey et al. 1998), and `list?` accepts proper lists only (Racket keeps it effectively constant-time by caching) (Flatt and PLT 2026). ISO Prolog (ISO/IEC 1995) has no name for such terms at all. Its built-ins raise `type_error(list, …)` for an argument that “is neither a partial list nor a list” (Hodgson 1999), where a partial list ends in an unbound variable. SWI-Prolog’s `is_list/1` accepts proper lists only. However, before version 5.0.1 it checked only the outermost constructor (Wielemaker 2026). Erlang’s reference manual first says that the tail of a list “is also a list”, and then that “it is allowed to have a list where the tail is not a list, for example, `[a|b]`. However, this type of list is of little practical use” (Ericsson AB 2026e). In CDuce (Benzaken et al. 2003; “CDuce Set Theoretic Type Checker” 2025), a sequence is a chain of pairs that ends in the atom `` `nil``. Thus, `[1 2]` and ``(1, (2, `nil))`` are the same value, and a chain that ends in another value is simply a nested pair (Section 4.1).

The table shows the aspect in which Erlang differs from the other four languages. It is the only one in which the word *list* means two different things. The guard `is_list/1` accepts any cons cell or `[]`, like Common Lisp’s `listp`, while the type `list()` contains proper lists only, like *list* in Scheme and Prolog. Elixir shares Erlang’s lists and the split.

### 4.1 Other type systems for lists

Of the five languages of Table 5, Erlang is the only one with a standard type for chains with a given element and termination type, `maybe_improper_list(T, U)`. Common Lisp’s type specifier `(cons T U)` describes a single cons cell, and its derived types cannot be recursive (Pitman 2005c; 2005d). Scheme and Prolog have no type language.

*Typed Racket.* Typed Racket (Tobin-Hochstadt and Felleisen 2008; PLT 2026) has `(Pairof s t)`, homogeneous lists `(Listof t)`, fixed-length lists `(List t ...)`, and `(List* t t1 ... s)`, which is `(Pairof t (List* t1 ... s))`, with `(List* s)` equivalent to `s` itself.

Table 5: Vocabulary for cons chains in five languages.

| A chain that … | Common Lisp | Scheme / Racket | Prolog (ISO) | CDuce | Erlang |
|---|---|---|---|---|---|
| is a cons cell or the empty list | list (`listp`) | pair or null | — | pair or `` `nil `` | accepted by `is_list/1` |
| ends in the empty list | proper list | list (`list?`) | list (`is_list/1` in SWI-Prolog) | sequence | proper list, type `list()` |
| ends in another value | dotted list | improper list | not a list (`type_error`) | a nested pair, not a sequence | improper list |
| ends in an unbound variable | — | — | partial list | — | — |
| does not end | circular list | circular list | cyclic term (SWI-Prolog) | impossible | impossible |

Improper lists are typed with `Pairof` and `List*`. The type of all chains over a given element and termination type can be defined recursively over `Pairof`. The zero-length case of `List*` is the convention of etylizer's helper `improper_list`$(t, u)$, whose use for `maybe_improper_list` caused the bugs of the first cons encoding (Appendix A.6).

*CDuce.* The encoding of Section 3 is the standard encoding of lists as recursive types over a product, and CDuce is its closest precedent, since it uses the same encoding for sequences in a set-theoretic type system. In CDuce (Benzaken et al. 2003; "CDuce Set Theoretic Type Checker" 2025), regular expression types in the style of XDuce (Hosoya and Pierce 2003), such as `[ Int String* ]`, CDuce therefore gives the same answers as etylizer, and offers in addition regular expression types and patterns as syntax, and samples in its error messages.

The differences lie mainly in the language that is checked. CDuce identifies lists with nested pairs and the empty list with an atom. Erlang's values do not allow this, since `{1, {2, []}}` is not `[1, 2]`. Therefore, etylizer keeps cons cells in a component of their own, separate from tuples, and `[]` as a constant of its own. CDuce's regular expression types denote `` `nil ``-terminated sequences only. A tail type is intersected with the sequence types. For example, `` [ Int ; `foo ] `` is empty, and a chain ending in `` `foo `` needs an explicit recursive pair type such as `` U = `foo | (Int, U) ``. Erlang's type language, by contrast, has list types with a termination, the guard `is_list/1` and `iolist()`, whose meaning had to be recovered first (Section 2). Finally, etylizer checks existing Erlang code against existing specifications, infers types for functions without one and supports gradual typing, whereas a CDuce program annotates every function.

## 5 Conclusion

todo

## References

"CDuce Set Theoretic Type Checker." 2025. https://gitlab.math.univ-paris-diderot.fr/cduce/cduce/.

"Types and Function Specifications." 2025. https://www.erlang.org/doc/system/typespec.html.

Armstrong, Joe. 2007. "A History of Erlang." In *Proceedings of the Third ACM SIGPLAN History of Programming Languages Conference (HOPL-III), San Diego, California, USA, 9-10 June 2007*, edited by Barbara G. Ryder and Brent Hailpern, *Proceedings of the Third ACM SIGPLAN History of Programming Languages Conference (HOPL-III), San Diego, California, USA, 9-10 June 2007.* ACM. https://doi.org/10.1145/1238844.1238850.

Benzaken, Véronique, Giuseppe Castagna, and Alain Frisch. 2003. "Cduce: An XML-Centric General-Purpose Language." In *Proceedings of the Eighth ACM SIGPLAN International Conference on Functional Programming, ICFP 2003, Uppsala, Sweden, August 25-29, 2003*, edited by Colin Runciman and Olin Shivers, *Proceedings of the Eighth ACM SIGPLAN International Conference* *on Functional Programming, ICFP 2003, Uppsala, Sweden, August 25-29, 2003.* ACM. https://doi.org/10.1145/944705.944711.

Castagna, Giuseppe, Kim Nguyen, Zhiwu Xu, and Pietro Abate. 2015. "Polymorphic Functions with Set-Theoretic Types: Part 2: Local Type Inference and Type Reconstruction." In *Proceedings of the 42nd Annual ACM SIGPLAN-SIGACT Symposium on Principles of Programming Languages, POPL 2015, Mumbai, India, January 15-17, 2015*, edited by Sriram K. Rajamani and David Walker, *Proceedings of the 42nd Annual ACM SIGPLAN-SIGACT Symposium on Principles of Programming Languages, POPL 2015, Mumbai, India, January 15-17, 2015.* ACM. https://doi.org/10.1145/2676726.2676991.

The Elixir Team. 2026. "Elixir V1.20: List.". https://hexdocs.pm/elixir/List.html.

Ericsson AB. 2026e. "Erlang Reference Manual: Data Types.". https://www.erlang.org/doc/system/data_types.html.

Ericsson AB. 2026c. "Erlang Reference Manual: Expressions.". https://www.erlang.org/doc/system/expressions.html.

Ericsson AB. 2026b. "Erlang/OTP 28.5 Source Code.". https://github.com/erlang/otp/tree/OTP-28.5.

Ericsson AB. 2026a. "ERTS Reference Manual: The erlang Module, Types.". https://www.erlang.org/doc/apps/erts/erlang.html#types.

Ericsson AB. 2026d. "ERTS User's Guide: The Abstract Format.". https://www.erlang.org/doc/apps/erts/absform.html.

Flatt, Matthew, and PLT. 2026. "The Racket Reference: Pairs and Lists.". https://docs.racket-lang.org/reference/pairs.html.

Frisch, Alain, Giuseppe Castagna, and Véronique Benzaken. 2008. "Semantic Subtyping: Dealing Set-Theoretically with Function, Union, Intersection, And Negation Types." *J. ACM* 55 (4): 19:1–19:64. https://doi.org/10.1145/1391289.1391293.

The Gleam Team. 2026. "The Gleam Language Tour.". https://tour.gleam.run/.

Hodgson, J. P. E. 1999. "Prolog: The ISO Directives, Control Constructs and Builtins.". http://www.deransart.fr/prolog/bips.html.

Högberg, John, and Ilya Klyuchnikov. 2024. "EEP 71: Clarify the Type Documentation, Especially Regarding Type Variables.". https://www.erlang.org/eeps/eep-0071.

Hosoya, Haruo, and Benjamin C. Pierce. 2003. "Xduce: A Statically Typed XML Processing Language." *ACM Transactions on Internet Technology (TOIT)* 3 (2): 117–48.

ISO/IEC. 1995. *Information Technology – Programming Languages – Prolog – Part 1: General Core.*

Kelsey, Richard, William Clinger, and Jonathan Rees. 1998. "Revised^5 Report on the Algorithmic Language Scheme". *ACM SIGPLAN Notices* 33 (9): 26–76. https://doi.org/10.1145/290229.290234.

Kiselyov, Oleg, Ralf Lämmel, and Keean Schupke. 2004. "Strongly Typed Heterogeneous Collections." In "Proceedings of the 2004 ACM SIGPLAN Workshop on Haskell." Special issue, *Proceedings of the 2004 ACM SIGPLAN Workshop on Haskell* (New York, NY, USA), 96–107. https://doi.org/10.1145/1017472.1017488.

Lindahl, Tobias, and Kostis Sagonas. 2007. "EEP 8: Types and Function Specifications.". https://www.erlang.org/eeps/eep-0008.

McCarthy, John. 1960. "Recursive Functions of Symbolic Expressions and Their Computation by Machine, Part I." *Communications of the ACM* 3 (4): 184–95. https://doi.org/10.1145/367177.367199.

Norell, Ulf. 2018. "What Is the Semantics of Improper Lists Anyway?.". https://github.com/josefs/Gradualizer/issues/110.

Pitman, Kent M. 2005a. "Common Lisp HyperSpec, Glossary.". https://www.lispworks.com/documentation/HyperSpec/Body/26_glo_d.htm.

Pitman, Kent M. 2005d. “Common Lisp HyperSpec, Macro DEFTYPE.”. https://www.lispworks.com/documentation/HyperSpec/Body/m_deftp.htm.

Pitman, Kent M. 2005c. “Common Lisp HyperSpec, System Class CONS.”. https://www.lispworks.com/documentation/HyperSpec/Body/t_cons.htm.

Pitman, Kent M. 2005b. “Common Lisp HyperSpec, System Class LIST.”. https://www.lispworks.com/documentation/HyperSpec/Body/t_list.htm.

PLT. 2026. “The Typed Racket Reference: Types.”. https://docs.racket-lang.org/ts-reference/type-ref.html.

Qqwy. 2022a. “Some Odd Typespecs That Use Maybe_improper_list Include [] As Possible `improper' Terminator.”. https://github.com/elixir-lang/elixir/issues/11886.

Qqwy. 2022b. “The Typespec of Iolist() Mentions [] As Possible `improper' Terminator.”. https://github.com/erlang/otp/issues/6038.

Schimpf, Albert. 2026. “Improper Lists.”. https://github.com/etylizer/etylizer/pull/363.

Schimpf, Albert, Stefan Wehr, and Annette Bieniusa. 2022. “Set-Theoretic Types for Erlang.” In “Proceedings of the 34th Symposium on Implementation and Application of Functional Languages, IFL 2022, Copenhagen, Denmark, 31 August 2022- 2 September 2022.” Special issue, *Proceedings of the 34th Symposium on Implementation and Application of Functional Languages, IFL 2022, Copenhagen, Denmark, 31 August 2022- 2 September 2022*, 4:1–4:14. https://doi.org/10.1145/3587216.3587220.

Schimpf, Albert, Stefan Wehr, and Annette Bieniusa. 2026. “Set-Theoretic Types for Erlang: Theory, Implementation, And Evaluation.”. https://doi.org/10.48550/arXiv.2603.22032.

Söderqvist, Viktor. 2020. “The Meaning of the Improper List Types in Erlang.”. https://github.com/zuiderkwast/improper.

Stack Overflow. 2009. “Functional Programming: What Is an ``improper List''?.”. https://stackoverflow.com/questions/1919097/functional-programming-what-is-an-improper-list.

Tobin-Hochstadt, Sam, and Matthias Felleisen. 2008. “The Design and Implementation of Typed Scheme.” In “Proceedings of POPL.” Special issue, *Proceedings of POPL*, 395–406. https://doi.org/10.1145/1328438.1328486.

Wielemaker, Jan. 2026. “SWI-Prolog Reference Manual: Is_list/1.”. https://www.swi-prolog.org/pldoc/doc_for?object=is_list/1.

## A Previous Approaches in etylizer

This appendix describes etylizer’s earlier approaches to lists. Three were unsound, others were sound but imprecise, and the treatment of improper lists combined both kinds of error.

### A.1 Lists as pairs

From 2022 to 2025, a list atom $\langle e, u\rangle$ stood for the non-empty chains whose elements are all in $e$ and whose final tail is in $u$:

$$\mathcal{L}(e,u) = \{\, [v_1, \dots, v_n \mid w] \mid n \geq 1,\ v_i \in [\![e]\!],\ w \in [\![u]\!] \,\},$$

where $[v_1, \dots, v_n \mid w]$ abbreviates $[v_1 \mid [v_2 \mid \dots [v_n \mid w] \dots]]$. The empty list was a separate constant, and $\texttt{list}(t) = \langle t, \texttt{[]}\rangle \vee \texttt{[]}$ and $\texttt{nonempty_list}(t) = \langle t, \texttt{[]}\rangle$. Emptiness was decided as for products (Frisch et al. 2008): an intersection $\bigwedge_{i\in P}\langle e_i, u_i\rangle \wedge \bigwedge_{j\in N} \neg\langle e'_j, u'_j\rangle$ with $E = \bigwedge_{i\in P} e_i$ and $U = \bigwedge_{i\in P} u_i$ counted as empty if and only if

$$\forall N' \subseteq N.\quad E \leq \bigvee_{j\in N'} e'_j \quad \text{or} \quad U \leq \bigvee_{j\in N\setminus N'} u'_j. \tag{1}$$

The type library was rewritten in 2023 and the decision diagrams were reimplemented several times. However, none of these changes touched the meaning of a list atom.

### A.2 Unsound: the product reading

Equation (1) is exact for the interpretation $[\![\langle e, u\rangle]\!]_\times = [\![e]\!] \times [\![u]\!]$, but the intended interpretation is $\mathcal{L}(e, u)$, and $\mathcal{L}$ is not a product.

*Counterexample.* For $\texttt{list}(\texttt{a} \vee \texttt{b}) \leq \texttt{list}(\texttt{a}) \vee \texttt{list}(\texttt{b})$, the `[]` parts cancel, and what remains is the emptiness of $\langle \texttt{a} \vee \texttt{b}, \texttt{[]}\rangle \wedge \neg\langle \texttt{a}, \texttt{[]}\rangle \wedge \neg\langle \texttt{b}, \texttt{[]}\rangle$. In (1), $E = \texttt{a} \vee \texttt{b}$ and $U = \texttt{[]}$. For every $N'$ that omits one of the negated atoms, the second disjunct holds, because the omitted atom has termination `[]` and $U \leq \texttt{[]}$; for $N'$ containing both, $\texttt{a} \vee \texttt{b} \leq \texttt{a} \vee \texttt{b}$ holds. The intersection is wrongly declared empty and the subtyping holds, but $\texttt{[a, b]} \in \mathcal{L}(\texttt{a} \vee \texttt{b}, \texttt{[]})$ and $\texttt{[a, b]} \notin \mathcal{L}(\texttt{a}, \texttt{[]}) \cup \mathcal{L}(\texttt{b}, \texttt{[]})$. The last revision with the pair encoding accepts `widen/1` from Section 1.

*Why.* Equation (1) rests on the identity $(E \times U) \setminus (E' \times U') = ((E \setminus E') \times U) \cup (E \times (U \setminus U'))$. For lists, $\mathcal{L}(E, U) \setminus \mathcal{L}(E', U')$ also contains lists that mix elements of $E \cap E'$ with elements of $E \setminus E'$. These lie in neither $\mathcal{L}(E \setminus E', U)$ nor $\mathcal{L}(E, U \setminus U')$. The element type constrains every element, but one pair stands for the whole list, and "every element is in $\texttt{a} \vee \texttt{b}$" does not split into "every element is in `a`" or "every element is in `b`". Intersection is not exact either once termination types overlap with cons cells. For an atom `x`, $\texttt{[1, a | x]} \in \mathcal{L}(\texttt{integer}(), \texttt{any}()) \cap \mathcal{L}(\texttt{any}(), \texttt{atom}())$ but $\texttt{[1, a | x]} \notin \mathcal{L}(\texttt{integer}(), \texttt{atom}())$.

The pattern approximations of Appendix A.4 were consistent only with this model. The upper type of the pattern `[1, 2]` was $\texttt{nonempty_list}(1) \vee \texttt{nonempty_list}(2)$, which does not contain `[1, 2]`. However, the pair model identifies it with $\texttt{nonempty_list}(1 \vee 2)$, which does. The pair *representation* is not the problem. Dialyzer and Elixir also describe lists by element and termination types, but neither decides emptiness as for products (Section 4).

### A.3 Unsound: the first constraint for cons expressions

Until late 2024, an expression $\texttt{[}e_1 \texttt{ | } e_2\texttt{]}$ with expected type $\tau$ produced the constraint $\texttt{list}(\alpha) \leq \tau$ with $e_1 : \alpha$ and $e_2 : \tau$, giving the tail the type expected for the whole list. If $\tau$ contains non-list values, the tail can be one of them. The revision before the change (`4ca3d51`, 2024-11-20) accepts

```
-spec f() -> list(integer()) | atom().
f() -> [1 | foo].
```

although `[1 | foo]` is neither a list of integers nor an atom. By contrast, it rejects the same body for the specification `list(integer())`.

### A.4 Unsound and imprecise: pattern approximations

A list pattern has no exact type in the pair model. Values matching `[1 | _]` have type `nonempty_list()`, but not every such value matches, and no type describes the lists of at least two elements that `[_, _ | _]` matches. A change started in 2022 and merged in 2023 gave each pattern $p$ two approximations, an upper type $\lceil p \rceil$, which every value matching $p$ has, and a lower type $\lfloor p \rfloor$, every value of which matches $p$:

$$\lceil \texttt{[1 | _]} \rceil = \texttt{nonempty_list}(\texttt{any}()) \qquad \lfloor \texttt{[1 | _]} \rfloor = \texttt{nonempty_list}(1)$$

$$\lceil \texttt{[_, _ | _]} \rceil = \texttt{nonempty_list}(\texttt{any}()) \lfloor \texttt{[_, _ | _]} \rfloor = \texttt{none}()$$

The cons case chose between these by ad-hoc subtyping tests on the tail type, and when the tail pattern was a variable or a wildcard it returned the proper list type $\texttt{nonempty_list}(\texttt{any}())$ as the upper type. That is unsound, because `[1 | foo]` matches `[E | Es]` but is not in $\texttt{nonempty_list}(\texttt{any}())$. The checker therefore assumed that an improper list never matches the first clause of `is_set/1` and falls through to the clause `_ -> false`. With the helper type `(list(), term()) -> boolean()`, the tail `Es` got a proper list type and the program was accepted.

On the other side, a `case` expression counted as exhaustive if the scrutinee type was a subtype of the union of the lower types. With these approximations the redundant third clause of the first `case` expression in Section 1 cannot be detected, which the developers recorded as a known limi-tation. Moreover, since the lower type of `[X | _]` was the proper list type `nonempty_list(any())`, improper lists were never covered.

The approximations also raised false alarms, which are worse than missed redundancy. Every string pattern had the type `string()`. As a result, the last revision with the pair encoding reports the live third clause of the second `case` expression in Section 1 with "this branch never matches". And since the lower types of `[_]` and `[_, _ | _]` were `none()`, it rejects the exhaustive match

```
case L of [] -> 0; [_] -> 1; [_, _ | _] -> 2 end
```

with "not all cases are covered". The cons encoding accepts both.

### A.5 Sound but imprecise

*No lengths, no positions.* A pair records what each element can be, not how many elements there are or which element is where. `[1, 2]`, `[2, 1]` and `[1, 1, 2]` all have type `nonempty_list`$(1 \lor 2)$, and string literals had type `nonempty_string()`.

*Constructing lists.* From late 2024, the tail of `[`$e_1$ `|` $e_2$`]` had to have a type `list`$(\beta)$, and the result type was `nonempty_list`$(\alpha \lor \beta)$. This is sound and keeps the result non-empty, but forces the tail to be a proper list. Consequently, `[1 | foo]` has no type at all.

### A.6 Improper lists before 2026

Table 6 shows the encodings of improper lists over time.

*Pairs.* The pair encoding translated `maybe_improper_list(`$t$`,`$u$`)` as $\langle t, u \rangle$, which lacks `[]`, rejecting valid programs. Its translation of `nonempty_improper_list(`$t$`,`$u$`)` as `list`$(t) \land \neg u$ is correct only for $u = $ `[]`; for other termination types it ignores $u$ and even contains `[]`, which is unsound. So is refining `X` to proper lists after `is_list(X)`, because the guard also succeeds for `[1 | 2]`.

*Cons, first cut.* The same translation, read through the helper `improper_list`$(t, u)$, includes chains of length zero, that is, $u$ itself. Hence

$$\texttt{maybe_improper_list()} = \mu X.\, \texttt{any}() \lor \texttt{cons}(\texttt{any}(), X) = \texttt{any}(),$$

and a bare binary was an `iolist()`, both unsound, while `maybe_improper_list`$(t, u)$ still lacked `[]` unless $u$ contained it. The printer showed the top list type as `improper_list`$(\texttt{any}(), \texttt{any}())$, which parsed back as `any`$()$.

Table 6: Encodings of improper lists over time. MIL$(t, u)$ and NEIL$(t, u)$ abbreviate `maybe_improper_list(`$t$`,`$u$`)` and `nonempty_improper_list(`$t$`,`$u$`)`, NEMIL() is `nonempty_maybe_improper_list()`, and IL$(t, u)$ is the helper $\mu X.\, u \lor \texttt{cons}(t, X)$. Entries marked † disagree with the semantics of Section 2. Before late 2024, MIL(), NEMIL() and `iolist()` were unsupported.

| | Pairs (2022–2025) | Cons, first cut (2025–2026) | Current (since June 2026) |
|---|---|---|---|
| MIL$(t, u)$ | † $\langle t, u \rangle$, no `[]` | † IL$(t, u)$, contains $u$ | `[]` $\lor$ NEIL$(t, u)$ |
| NEIL$(t, u)$ | † `list`$(t) \land \neg u$ | `cons`$(t, \text{IL}(t, u))$ | same |
| MIL() | † $\langle$`any()`, `any()`$\rangle$ | † `any()` | `[]` $\lor$ `cons(any(), any())` |
| NEMIL() | † `nonempty_list()` | † `nonempty_list()` | `cons(any(), any())` |
| `iolist()` | † no top-level `[]` | † contains `binary()` | as in Section 3.2 |
| `is_list(X)` establishes | † `list()` | † `list()` | MIL() |
| `[`$e_1$ `|` $e_2$`]` requires | † proper tail | any tail | any tail |

### A.7 Why etylizer got it wrong

etylizer's translation of improper list types was wrong from 2022 until 2026, in two different ways (Appendix A.6). In retrospect, five factors contributed.

- *No written semantics.* The documentation defers to "the set of Erlang terms one would expect", and the three OTP components that fix a meaning do so in code, for their own purposes, and disagree on the empty list.
- *Pair-shaped vocabulary.* "Contents" and "termination", and Dialyzer's representation, describe a list by two types. etylizer's first encoding did the same (Appendix A.1).
- *A deliberate simplification.* The 2022 translation, citing a long discussion of improper lists in Gradualizer's issue tracker, chose "to keep things simple" not to distinguish improper from maybe-improper lists. It mapped `maybe_improper_list(T, U)` to a non-empty list type, losing `[]`, and left the zero-argument `nonempty_maybe_improper_list()` with the comment "TODO what is a maybe improper list vs improper list?".
- *A translation that outlived its target.* When lists became cons cells in 2025, the translation of `maybe_improper_list(T, U)` was kept and now produced the helper `improper_list`$(t, u)$, which contains chains of length zero, that is, $u$ itself. Typed Racket's `List*` has the same convention (Section 4). This convention is right for a helper and wrong for `maybe_improper_list`.
- *An overloaded word.* In a guard, a list is a cons cell or `[]`, whereas the type `list()` contains proper lists only. etylizer read `is_list(X)` as establishing `list()`, which Dialyzer uses only to narrow later clauses. Typing the guarded branch needs `maybe_improper_list()`. Gradualizer and eqWAlizer still make the same choice.